\documentclass[prb,twocolumn,showpacs,preprintnumbers,amsmath,amssymb]{revtex4-1}
\usepackage{ifpdf}
 \newif\ifpdf
\ifx\pdfoutput\undefined
   \pdffalse
\else
   \pdfoutput=1
   \pdftrue
\fi
\ifpdf
   \usepackage{graphicx}
   \usepackage{epstopdf}
   \usepackage{color}
\else
   \usepackage{graphicx}
\fi

\usepackage{srctex}
\usepackage{hyperref}

\usepackage{amsmath,amssymb}

\usepackage{epsfig}

\usepackage{epsfig}

\begin{document}

\title{Cooling rate dependence of simulated ${\rm Cu_{64.5}Zr_{35.5}}$ metallic glass structure}

\author{R.E. Ryltsev}
\affiliation{Institute of Metallurgy, Ural Branch of Russian Academy of Sciences, 620016, 101 Amundsen str., Ekaterinburg, Russia}
\affiliation{Ural Federal University, 620002, 19 Mira str., Ekaterinburg, Russia}
\affiliation{L.D. Landau Institute for Theoretical Physics, Russian Academy of Sciences, 119334, 2 Kosygina str., Moscow, Russia}

\author{B.A. Klumov}
\affiliation{Aix-Marseille-Universit\'{e}, CNRS, Laboratoire PIIM, UMR 7345, 13397 Marseille cedex 20, France}
\affiliation{High Temperature Institute, Russian Academy of Sciences, 125412, 13/2 Izhorskaya str., Moscow, Russia}
\affiliation{L.D. Landau Institute for Theoretical Physics, Russian Academy of Sciences, 119334, 2 Kosygina str., Moscow, Russia}

\author{N.M. Chtchelkatchev}
\affiliation{L.D. Landau Institute for Theoretical Physics, Russian Academy of Sciences, 119334, 2 Kosygina str., Moscow, Russia}
\affiliation{Moscow Institute of Physics and Technology, 141700б 9 Institutskiy per., Dolgoprudny, Moscow Region, Russia}
\affiliation{Institute of Metallurgy, Ural Branch of Russian Academy of Sciences, 620016, 101 Amundsen str., Ekaterinburg, Russia}

\author{K.Yu. Shunyaev}
\affiliation{Institute of Metallurgy, Ural Branch of Russian Academy of Sciences, 620016, 101 Amundsen str., Ekaterinburg, Russia}
\affiliation{Ural Federal University, 620002, 19 Mira str., Ekaterinburg, Russia}


\begin{abstract}
Using molecular dynamics simulations with embedded atom model potential, we study structural evolution of ${\rm Cu_{64.5}Zr_{35.5}}$ alloy during the cooling in a wide range of cooling rates $\gamma\in(1.5\cdot 10^{9},10^{13})$ K/s. Investigating short- and medium-range order, we show that structure of ${\rm Cu_{64.5}Zr_{35.5}}$ metallic glass essentially depends on cooling rate. In particular, a decrease of the cooling rate leads to a increase of abundances of both the icosahedral-like clusters and Frank-Kasper Z16 polyhedra. The amounts of these clusters in the glassy state drastically increase at the $\gamma_{\rm min}=1.5\cdot 10^{9}$ K/s. Analysing the structure of the glass at $\gamma_{\rm min}$, we observe the formation of nano-sized crystalline grain of ${\rm Cu_2Zr}$ intermetallic compound with the structure of ${\rm Cu_2Mg}$ Laves phase. The structure of this compound is isomorphous with that for ${\rm Cu_5Zr}$ intermetallic compound.  Both crystal lattices consist of two type of clusters: Cu-centered 13-atom icosahedral-like cluster and Zr-centered 17-atom Frank-Kasper polyhedron Z16. That suggests the same structural motifs for the metallic glass and intermetallic compounds and explains the drastic increase of the abundances of these clusters observed at $\gamma_{\rm min}$.
\end{abstract}
\pacs{61.20.Gy, 61.20.Ne, 64.60.Kw}

\maketitle
\section{Introduction}

Bulk metallic glasses (BMGs) are in focus of intense experimental and theoretical research last decade due to their extraordinary physical properties with regards to their crystalline counterparts and high potential for advanced applications \cite{suryanarayana2010bulk,Trexler2010ProgMatSci,Wang2004MatSciEngR}.  BMGs are bulk metallic alloys with only short- and medium-range order structure that makes them outstanding compared to crystalline metal alloys where the contact areas between the crystalline domains are usually weak points. However, the difficulty to construct a BMG with high enough glass forming ability prevents widespread applications of these new materials.

Recently, considerable efforts have been focused on studying of Cu-Zr alloys. It is one of the most extensively studied binary metallic systems, mainly due to its ability to form BMGs  \cite{Xu2004ActMat,Wang2004AppPhysLett,Wang2005JMatRes}. The compositions of BMG formation in Cu-Zr alloys are located in narrow (so-called pinpoint) concentration intervals \cite{Wang2004AppPhysLett,Li2008Science,Yang2012PRL}.

A general idea that local icosahedral ordering plays the key role in glass formation has been experimentally suggested as common feature of metallic glasses \cite{Moritz2002PRL}. For Cu-Zr system, this idea has been strongly supported by molecular dynamics simulations \cite{Li2009PRB,Peng2010ApplPhysLett,Soklaski2013PRB,Wu2013PRB,Wen2013JNonCrystSol,Wang2015JPhysChemA}.

Important question is the nature of glass forming ability of Cu-Zr alloys. The main conclusion of the extensive researches is the idea that local structure of glass-forming alloys has great influence on dynamical properties and glass-forming ability, see Ref.~\cite{Royall2015PhysRep} for review. A general idea that local icosahedral ordering plays the key role in glass formation has been experimentally suggested as common feature of metallic alloys \cite{Moritz2002PRL}. For Cu-Zr system, this idea has been strongly supported by molecular dynamics simulations \cite{Li2009PRB,Peng2010ApplPhysLett,Soklaski2013PRB,Wu2013PRB,Wen2013JNonCrystSol,Wang2015JPhysChemA}. It has been shown that icosahedral clusters tend to connect each other under cooling and form percolating network as the system approaches glass transition \cite{Li2009PRB,Soklaski2013PRB}. Besides, the correlation between icosahedral ordering and glassy dynamics has been ascertained \cite{Wu2013PRB,Zhang2015PRB}.

Despite of wide acceptance of the icosahedral order, alternative clusters have been considered to be important structural elements in Cu-Zr metallic glasses \cite{Li2009PRB,Peng2010ApplPhysLett,Ding2014ActMat,Zemp2014PRB,Ward2013PRB,Fang2011SciRep,Sun2016SciRep}. It has been particulary shown that, except Cu-centered icosahedra, Zr-centered Frank-Kasper polyhedra play important role in structural formation and dynamical slowing down \cite{Li2009PRB,Peng2010ApplPhysLett,Ding2014ActMat,Zemp2014PRB,Sun2016SciRep}. Models of ideally packed clusters \cite{Ward2013PRB} and Bergamon-type medium-range order~\cite{Fang2011SciRep} have been also proposed as possible candidates to describe structure of Cu-Zr glasses. So the problem of recognition of structural elements responsible for the glass formation in Cu-Zr alloys is still unsolved.

The possible way to determine local structural elements of metallic glasses is turning to those in crystalline state. The structure of rapidly quenched glass-forming alloys  often includes nano- and micro-sized grains of intermetallic compounds \cite{Hirata2007Ultramicroscopy, Chen2006ApplPhyLett} whose elements may serve as structural motifs for glassy state. According to equilibrium phase diagram \cite{Okamoto2012JPhasEq}, Cu-Zr system has a lot of stable intermetallic compounds and even more of them can form as metastable ones in non-equilibrium conditions \cite{Zhou2010ActaMater, Koval1992ScriptMet}. That suggests a lot of possible candidates to local structure elements. But cooling rates available in simulations are almost always too fast to form any crystalline grains.

It has been widely accepted that glass properties depend on the cooling rate $\gamma$ \cite{Zhang2014AppPhysLett,Zhang2015PRB_2,Yanjun2015TrNonFerMetChin,Vollmayr1996JCP}. In particular, the glass transition temperature $T_g$ varies with $\gamma$; for example, there is empirical formula $T_g(\gamma)\propto 1/({\rm{const}} - \ln \gamma)$ \cite{Vasin2011JStatMech}. So understanding nature of the glass-forming ability in BMG should be strongly correlated with  the cooling rate investigations. This is a challenge for numerical simulations usually limited by the microsecond timescale.

In this paper we systematically study the cooling rate dependence of structure of ${\rm Cu_{64.5}Zr_{35.5}}$ alloy focusing on local orientational order. Considering the wide range of cooling rates $\gamma\in(1.5\cdot 10^{9}, 10^{13})$~K/s, we show that the structure of the glassy state essentially depends on $\gamma$. At the lowest cooling rate used we observe formation of nano-sized crystalline grain ${\rm Cu_2Zr}$ whose structural elements are both icosahedra and Frank-Kasper clusters widely accepted as structural motifs for metallic glasses.

\section{Methods}

\subsection{Simulations details}

Classical molecular dynamics (MD) is the main theoretical tool to study properties of glasses because it makes it possible to overcome the problems of analytical description of non-ordered condensed matter systems \cite{Liu1987PRB,Dubinin2014RussianChemRev,Dubinin2014RusChemRev_Err,Dubinin2008JPCM} and allows studying microscopic structure and dynamics covering sufficiently large time and spatial scales.

For MD simulations, we used $\rm{LAMMPS}$ Molecular Dynamics Simulator \cite{Plimpton1995JCompPhys}. The system of $N=5000$ particles was simulated under periodic boundary conditions in Nose-Hoover NPT ensemble at $P=0$.  The MD time step was varied from 1 to 3 fs depending on system temperature. We checked that chosen MD step values provided good energy conservation at given thermodynamic conditions.

Initial configuration was prepared as hcp-lattices with random seeding of the species in the lattice sites. This configuration was melted and completely equilibrated at $T=1800$~K. Then the system was cooled from $T=1800$~K down to $T=300$~K with different cooling rates $\gamma=\Delta T/\Delta t$ in the interval of $\gamma\in(1.5\cdot 10^{9}, 10^{13})$ K/s.

As the model of interaction between alloy components, we use widely accepted embedded atom model potential~\cite{Daw1993MatSciRep} of  Finnis-Sinclair type~\cite{Finnis1984PhilMag} developed by Mendelev et.al.~\cite{Mendelev2009PhilMag,Mendelev2009PhilMag_2}. This potential was specially designed to describe liquid and glassy states of the Cu-Zr alloys.

\subsection{Methods of structure analysis}

To study the local orientational order, we use the method of bond order parameters (BOP) \cite{Steinhardt1981,Steinhardt1983} which is widely accepted in the context of condensed matter physics (e.g. \cite{Steinhardt1983,Mitus1982,Trus2000}), hard sphere and Lennard-Jones systems~\cite{TenWolde,ErringtonLJ,RT96,Troadec99,Luchnikov2002,Jin2010,KlumovLJ13a,KlumovLJ13b,Bar14,KLumovJPC14}, both bulky and confined complex plasmas~\cite{CPP04,RZ06,JETP08,JETPL09,PPCF09,KlumovPU10,Khrapak2012}, colloidal suspensions \cite{Gasser01,Kawasaki10}, patchy systems~\cite{Vasilyev2013,Vasilyev2015}, etc.

The method allows us to explicitly recognize atomic clusters of any symmetry \cite{KlumovPU10,KlumovPRB11,Hirata2013Science} and study their spatial distribution \cite{Ryltsev2013PRE}. Below, we will briefly consider the key points of the method.

Within the frameworks of the BOP method, the rotational invariants of rank $l$ of both second $q_l({\bf r}_i)$ and third $w_l({\bf r}_i)$ order are calculated for each particle $i$ located at the point ${\bf r}_i$ from the vectors (bonds) ${\bf r}_{ij}$ connecting its center with the centers of its $N_{\rm nn}({\bf r}_i)$ nearest neighboring particles:
\begin{equation}\label{ql}
q_l^2 ({\bf r}_i ) = \frac{{4\pi }}{{2l + 1}}\sum\limits_{m =  - l}^l {\left| {q_{lm} ({\bf r}_i )} \right|^2 }
\end{equation}
\begin{equation}\label{wl}
w_l ({\bf r}_i ) = \sum\limits_{\{ m_i \} } {\left[ \begin{array}{l}
 l\quad l\quad l \\
 m_1 m_2 m_3  \\
 \end{array} \right]} \,q_{lm_1 } ({\bf r}_i )q_{lm_2 } ({\bf r}_i )q_{lm_3 } ({\bf r}_i )
\end{equation}
where $q_{lm}({\bf r}_i) = N_{\rm nn}({\bf r}_i)^{-1} \sum_{j=1}^{N_{\rm nn}({\bf r}_i)} Y_{lm}(\varphi_{ij},\theta_{ij})$, $Y_{lm}$ are the spherical harmonics and $(\varphi_{ij},\theta_{ij})$ are polar and azimuthal angles of the vectors ${\bf r}_{ij} = {\bf r}_i - {\bf r}_j$ connecting centers of particle $i$ and $j$. In Eq.(\ref{wl}) $\left[ \begin{array}{l}
 l\quad l\quad l \\
 m_1 m_2 m_3  \\
\end{array} \right]$ are the Wigner 3$j$-symbols, and the summation in the latter expression is performed over all the indexes $m_i =-l,...,l$ satisfying the condition $m_1+m_2+m_3=0$.

The big advantage of invariants $q_l$ and $w_l$ is that they are uniquely determined for any polyhedron including the elements of any crystalline structure. By varying number of nearest neighbors $N_{\rm nn}$ and rank $l$ of BOP it is possible to identify clusters existing in the system. Among the parameters (\ref{ql}),(\ref{wl}), the $q_4$, $q_6$, $w_4$, $w_6$ are typically one of the most informative one so we use they in this work. In Tab.~\ref{tab1:inv} we present the values of these parameters for few different clusters.

 To identify the symmetry of the local clusters, we calculate the rotational invariants $q_l$ and $w_l$ for each atom using the fixed number of nearest neighbors  (e.g. $N_{\rm nn}=12$ for the closely packed structures like hcp, fcc, icosahedron and liquid-like $\rm CuZr$ system; to detect Frank-Kasper (FKs) polyhedrons (see insets in Fig.~\ref{fig4:kfs}) we use
$N_{\rm nn}=16,15,14$ for the Z16, Z15 and Z14, respectively). Atom whose coordinates in the space $(q_4, q_6, w_6)$ are sufficiently close to those for perfect structures is counted as fcc-like (hcp-like, icosahedral-like) etc.

\begin{table}[th]
\centering
\caption{Bond order parameters (BOPs) $q_l$ and $w_l$ ($l=4,~6$) of a few perfect clusters calculated via fixed number of nearest neighbors (NN): hexagonal close-packed (hcp), face centered cubic (fcc), icosahedron (ico), body centered cubic (bcc),  Frank-Kasper polyhedra Z14, Z15 and Z16. Additionally, mean BOPs for the LJ melt are indicated for the comparison}.
\begin{tabular}{|c|c|c|c|c|}
\hline 
lattice type & \quad $q_{4}$ & \quad $q_{6}$ & \quad $w_{4}$ & \quad $ w_{6}$
\\ \hline hcp (12 NN) & 0.097 & 0.485 & 0.134  & -0.012
\\ \hline fcc  (12 NN) & 0.19  & 0.575  & -0.159 &  -0.013
\\ \hline ico  (12 NN) & $1.4 \times 10^{-4}$ & 0.663 & -0.159  & -0.169
\\ \hline bcc ( 8 NN) & 0.5 & 0.628 & -0.159   & 0.013
\\ \hline bcc (14 NN) & 0.036 & 0.51 & 0.159   & 0.013
\\ \hline Z14 (14 NN) & 0.015 & 0.406 & 0.134   & -0.091
\\ \hline Z15 (15 NN) & 0.039 & 0.316 & 0.134   & -0.0405
\\ \hline Z16 (16 NN) & 0.0108 & 0.1924 & 0.1593   & 0.01316
\\ \hline LJ melt (12 NN) & $\approx $0.155  & $\approx $0.37  & $\approx $-0.023  &$\approx $-0.04
\\\hline 
\end{tabular}
\label{tab1:inv}
\end{table}

\section{Results}

\subsection{Influence of the cooling rate on system relaxation: continuous cooling vs sub-$T_g$ annealing}

\begin{figure*}
  \centering
  \includegraphics[width=\textwidth]{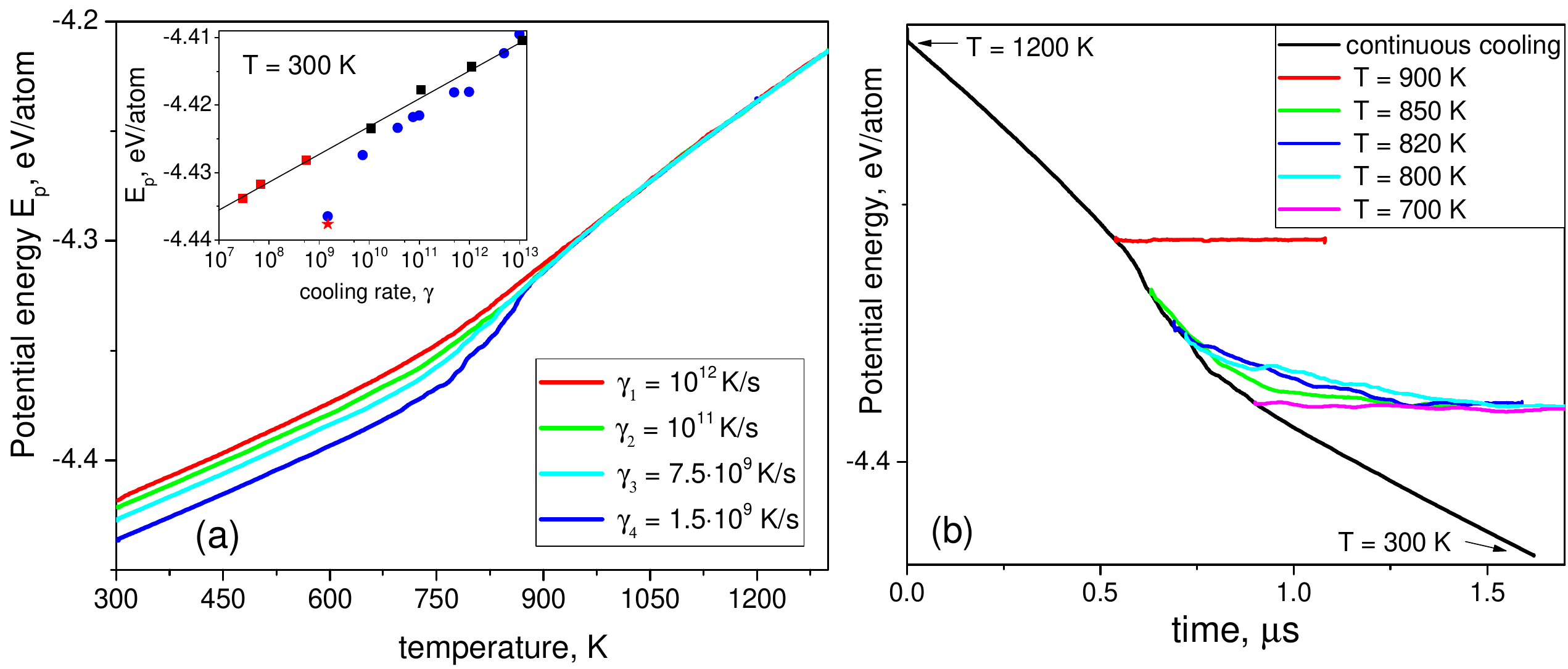}\\
  \caption{(Color online) (a) Temperature dependencies of system potential energy at different cooling rates. Inset shows dependence of room temperature value of potential energy on the cooling rate. (Blue) bullets represent values obtained at continuous cooling and (red) star is the one obtained at $\gamma=1.5\cdot 10^9$ K/s with additional sub-$T_g$ annealing at $T=700$. Squares and diamonds are the same value from \cite{Zhang2015PRB_2}. Here (black) squares are obtained at continuous cooling and (red) diamonds represent the results of sub-$T_g$ annealing. For the latter case effective cooling rates have been estimated by linear fit. The uncertainties of $E_p$ determined by averaging over time fluctuations are of the order of the point size. (b) Time dependencies of system potential energy during the continuous cooling and the annealing at different temperatures. In both figures, short-time fluctuations of $E_p$ have been smoothed by averaging over appropriate time window.}
  \label{fig1:Epot}
\end{figure*}

In Fig.~\ref{fig1:Epot}a we show smoothed temperature dependencies of system potential energy $E_{p}$ obtained under continuous cooling at different cooling rates. All the $E_p(T)$ curves demonstrate inflections indicating the liquid-glass transition. The glass transition temperature $T_g$ slightly decreases with the decrease of the cooling rate. We also see that a decrease of the cooling rate causes essential decrease of the potential energy at $T\lesssim T_g$. So the system is better equilibrated at lower $\gamma$.

Note that minimal cooling rates available in computer simulations are of the order of $10^9$~K/s. Cooling a system of $N\sim10^{4}$ particles from liquid to glassy state even at such $\gamma$ requires already about a month of calculations on supercomputer cluster and further essential reduction of $\gamma$ is hardly possible. The fact that $\gamma \sim 10^9$ K/s is still too fast in comparison to experimental cooling rates (which are of the order of $1-10^7$ K/s) leads researchers to look for other effective methods of simulating amorphization. Recently, the method of sub-$T_g$ annealing has been proposed for this purpose \cite{Zhang2014AppPhysLett}. It has been suggested that annealing the system at temperature which is slightly lower than the glass transition temperature $T_g$ is equivalent to the effective reducing of cooling rate. This method has been applied to ${\rm Cu_{64.5}Zr_{35.5}}$ alloy \cite{Zhang2015PRB_2} and the effective cooling rate obtained has been declared to be of the order of $10^7$ K/s (with minimal real cooling rate of the order of $10^{10}$ K/s).

Inspired by the papers mentioned above, we have performed additional annealing of the system at few different temperatures near $T_g$. As the initial configurations for annealing, the states obtained for correspondence temperatures under cooling at the lowest $\gamma_{\rm min}=1.5\cdot 10^9$ K/s were collected.  In Fig.~\ref{fig1:Epot}b, we demonstrate smoothed time dependencies of system potential energy $E_p(t)$ for both the continuous cooling at $\gamma_{\rm min}$ and the isothermal annealing at different temperatures. As seen from the picture, the average potential energy is a constant during the annealing at $T=900$ K. It reflects the fact that the continuous cooling at $\gamma_{\rm min}$ down to this temperature is a quasi-equilibrium process. The similar $E_p(t)$ dependence is observed during the annealing at $T=700$ K. But the reason is the opposite: the system is in glassy state and so the relaxation is frozen. But the annealing at $700<T<900$ K leads to lowering the potential energy that means the presence of system relaxation. Note that the lower the annealing temperature the lower the relaxation rate is observed. But, regardless of annealing temperature, the final value of the average potential energy at $t \to \infty$ is the same and so the system relaxes to similar states.

After annealing, the system was cooled down to $T=300$ K at the same $\gamma$.  In the inset of Fig.~\ref{fig1:Epot} we show the potential energy $E_p^{\rm(300)}$ of the final state at $T=300$ as the function of the cooling rate (blue bullets). The $E^{\rm(300)}_{p}(\gamma_{\rm min})$ value obtained with the additional annealing is presented by the red star. We see that this sub-$T_g$ annealing does not change the results essentially at least at used annealing times of the order of microseconds. The $E_p^{\rm(300)}$  obtained in Ref.~\onlinecite{Zhang2015PRB_2} at different (effective) cooling rates are also shown in the inset of Fig.~\ref{fig1:Epot}. The squares represent $E_p^{\rm(300)}$  values obtained under continuous cooling and the diamonds are the results of sub-$T_g$ annealing for different times (0.1, 0.5 and 2 $\mu {\rm s}$). For the latter case, the effective cooling rates have been estimated by logarithmic extrapolation (which is linear in logarithmic scale used in Fig.~\ref{fig1:Epot}). Comparison of the results reveal essential deviation of $E_p^{\rm(300)}(\gamma)$ dependence from logarithmic extrapolation at $\gamma < 10^{10}$ K/s. Below we will see that this deviation is caused by nanocrystallization which takes place at $\gamma=1.5\cdot 10^{9}$ K/s (see sec.~\ref{sec:lowest_v} and sec.~\ref{sec:disc1} for discussion).

\subsection{Two-point correlation functions}

\begin{figure}
  \centering
  \includegraphics[width=\columnwidth]{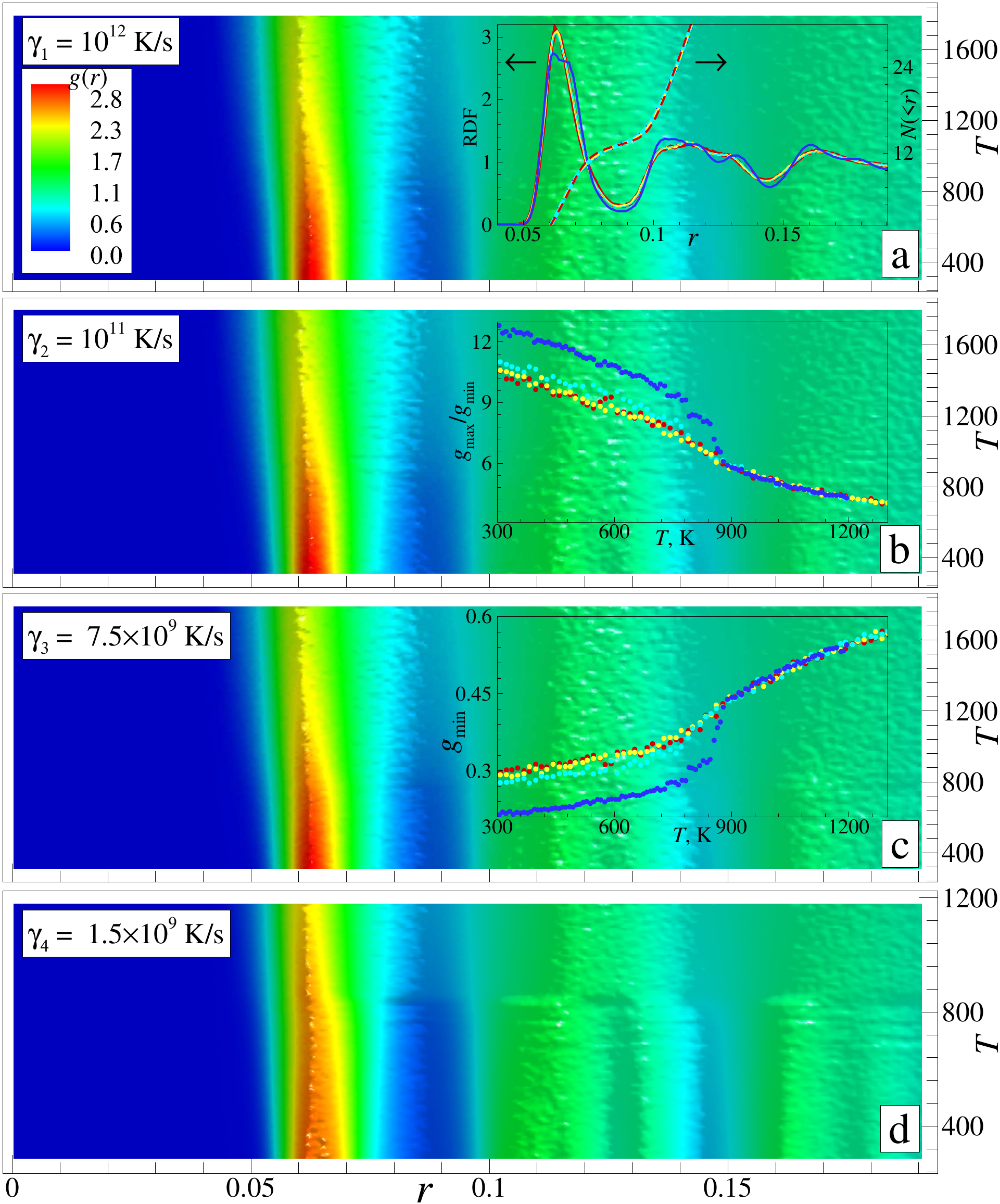}\\
  \caption{(Color online). Temperature evolution of the total radial distribution function $g(r)$ at different cooling rates $\gamma$ (indicated on the plot). The color on the $r-T$ plane represents the $g(r)$ values, see the scale in the panel (a). Splitting of the second peak of $g(r)$ which is usually referred as an indicator of the glass transition can be seen for all considered $\gamma$ close to $T_{\rm g} \simeq 850$ K; at $\gamma = 1.5\times 10^9$ K/s (panel d) the effect is the most pronounced. Insets at panels  present: (a) $g(r)$ for final glassy states at different $\gamma$ and associated cumulative functions $N(<r)$ showing closely packed neighboring atoms; (b),(c) show temperature dependencies of structure-sensitive indicators $g_{\rm max}/g_{\rm min}$ and  $g_{\rm min}$ at different cooling rates (see the text for explanation). For all the curves in insets:  $\gamma_4 = 1.5\times 10^9$ K/s (blue), $\gamma_3 = 7.5\times 10^9$ K/s (cyan), $\gamma_2 = 10^{11}$ K/s (yellow) and $\gamma_1 = 10^{12}$ K/s (red).}
  \label{fig1:rdf}
\end{figure}

Properties of the short- and medium-range translational order can be evaluated from the analysis of the total translational two-point correlation function - radial distribution function $g(r)$. Fig.~\ref{fig1:rdf} shows how the $g(r)$ of the ${\rm Cu_{64.5}Zr_{35.5}}$ system varies with the temperature $T$ at different cooling rates $\gamma$. Inset in Fig.~\ref{fig1:rdf}a presents total $g(r)$ for final glassy states and associated cumulative functions $N(<r)$ showing number of the nearest neighbors.  Splitting of the second peak of $g(r)$ which is usually referred as an indicator of the glass transition \cite{Finney1977Nature,Zallen2008book} can be seen for all considered $\gamma$ close to $T_{\rm g} \simeq 850$ K; at $\gamma = 1.5\times 10^9$ K/s (panel d) the effect is the most pronounced. Insets in Fig.~\ref{fig1:rdf}b,c show another structure-sensitive indicators $g_{\rm max}/g_{\rm min}$ and  $g_{\rm min}$, where $g_{\rm max}$ and $g_{\rm min}$ are respectively the values of $g(r)$ at the first maximum and the first non-zero minimum. These parameters have been proposed to be good indicators of structural changes taking place at melting and freezing \cite{Raveche1974JCP, KlumovLJ13a, KlumovLJ13b} and glass transition \cite{Wendt1978PRL}. Temperature dependencies of these indicators demonstrate pronounced inflections at the same temperatures as the inflections at $E_p(T)$ (Fig.~\ref{fig1:Epot}a) and the splitting of the second $g(r)$ peak that supports the cited estimation of $T_{\rm g}$.

\begin{figure}
  \centering
  \includegraphics[width=\columnwidth]{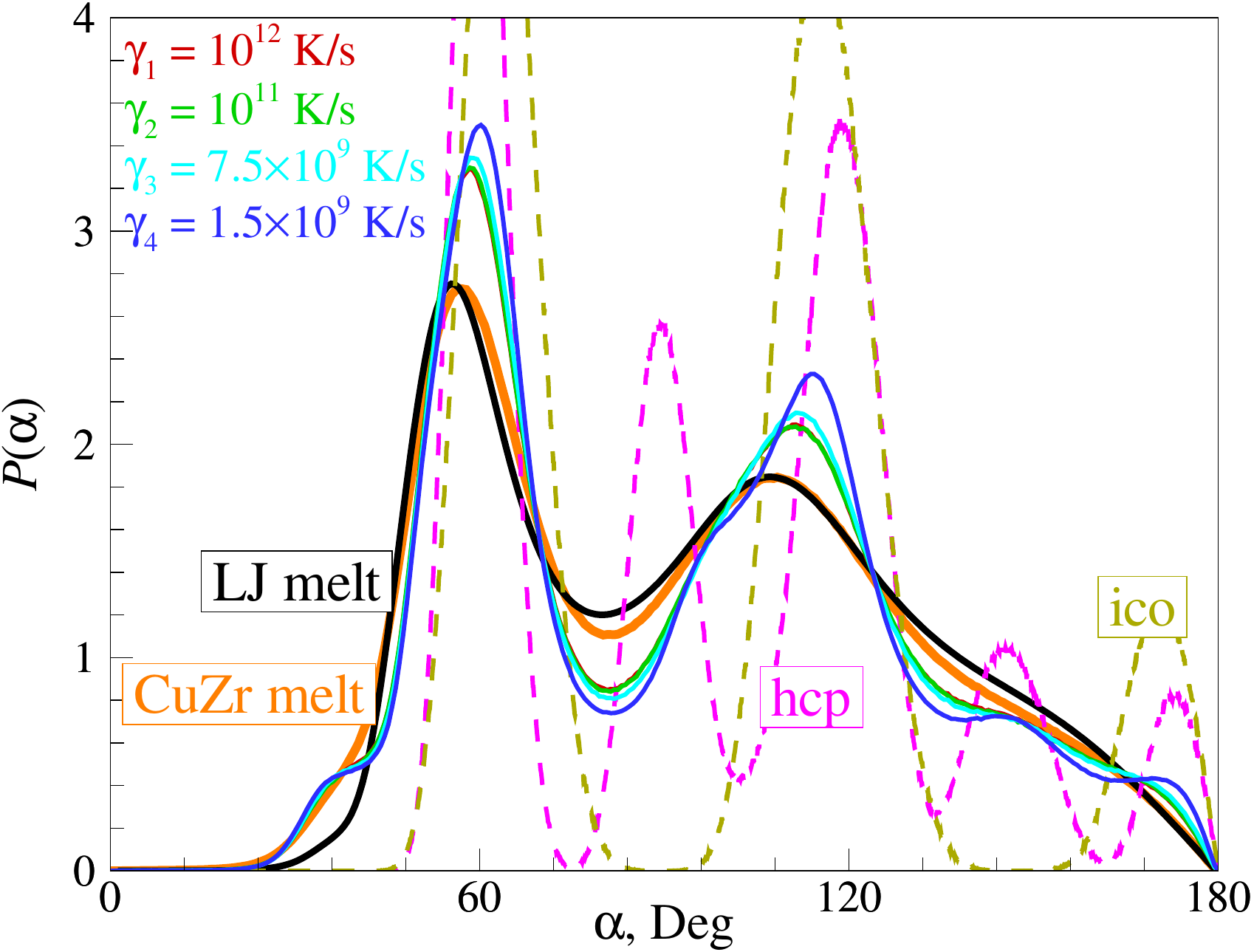}\\
  \caption{(Color online) Bond angle distribution function (BADF) for final (taken at $T \approx $ 300 K) glassy state of
  the ${\rm Cu_{64.5}Zr_{35.5}}$ system obtained at different cooling rates $\gamma_i$ (indicated on the plot) compared with those for hcp clusters (dashed magenta) and icosahedral clusters (dashed orange). Additionally, BADFs of melt for the Lennard-Jones (solid black) and the ${\rm Cu_{64.5}Zr_{35.5}}$ (solid orange) systems are plotted for the comparison.}
  \label{fig:angle_distr}
\end{figure}

Another important two-point correlation function, bond angle distribution function (BADF) $P(\alpha)$, measures the probability that two nearest neighbors and central atom form the angle $\alpha$. Fig.~\ref{fig:angle_distr} shows $P(\alpha)$ at different cooling rates $\gamma$ for the final glassy state of the system at temperature $T = 300$ K. We note that the BADFs reveal weak cooling rate dependence at $\gamma \ge 10^{10}$ K/s.  $P(\alpha)$ for weakly disturbed hcp and icosahedral clusters are plotted to find evidence of them in the glassy states of the system under consideration. Indeed the $P(\alpha)$ for glassy state demonstrate pronounced maxima which are close to those for icosahedral cluster. The correspondence between BADFs for the glass and a hcp cluster is not so clear. More detailed analysis will be performed below with using BOP method (see sec.\ref{sec:locorder}).

To analyse the structure of high-temperature liquid from which the glass has been prepared, we plot in Fig.~\ref{fig:angle_distr} BADFs for both ${\rm Cu_{64.5}Zr_{35.5}}$  melt at $T=1200$ K and Lennard-Jones liquid at $T^* \approx 1.5$, $\rho^* \approx 1$ (in reduced Lennard-Jones units \cite{Ryltsev2013PRE}); so the thermodynamic states of both systems correspond to a dense liquid close to the corresponding solid-liquid coexistence line. We see that BADFs of the systems under consideration are very close that suggests some similarities in between. The origin of that similarity can be explained by the existence of tetrahedral order which has been found in Lennard-Jones fluids \cite{Ryltsev2013PRE}. Our results suggests the formation of local tetrahedra might be the universal feature of close packed fluids.

\subsection{Local orientational order\label{sec:locorder}}

\begin{figure}
\centering
\includegraphics[width=\columnwidth]{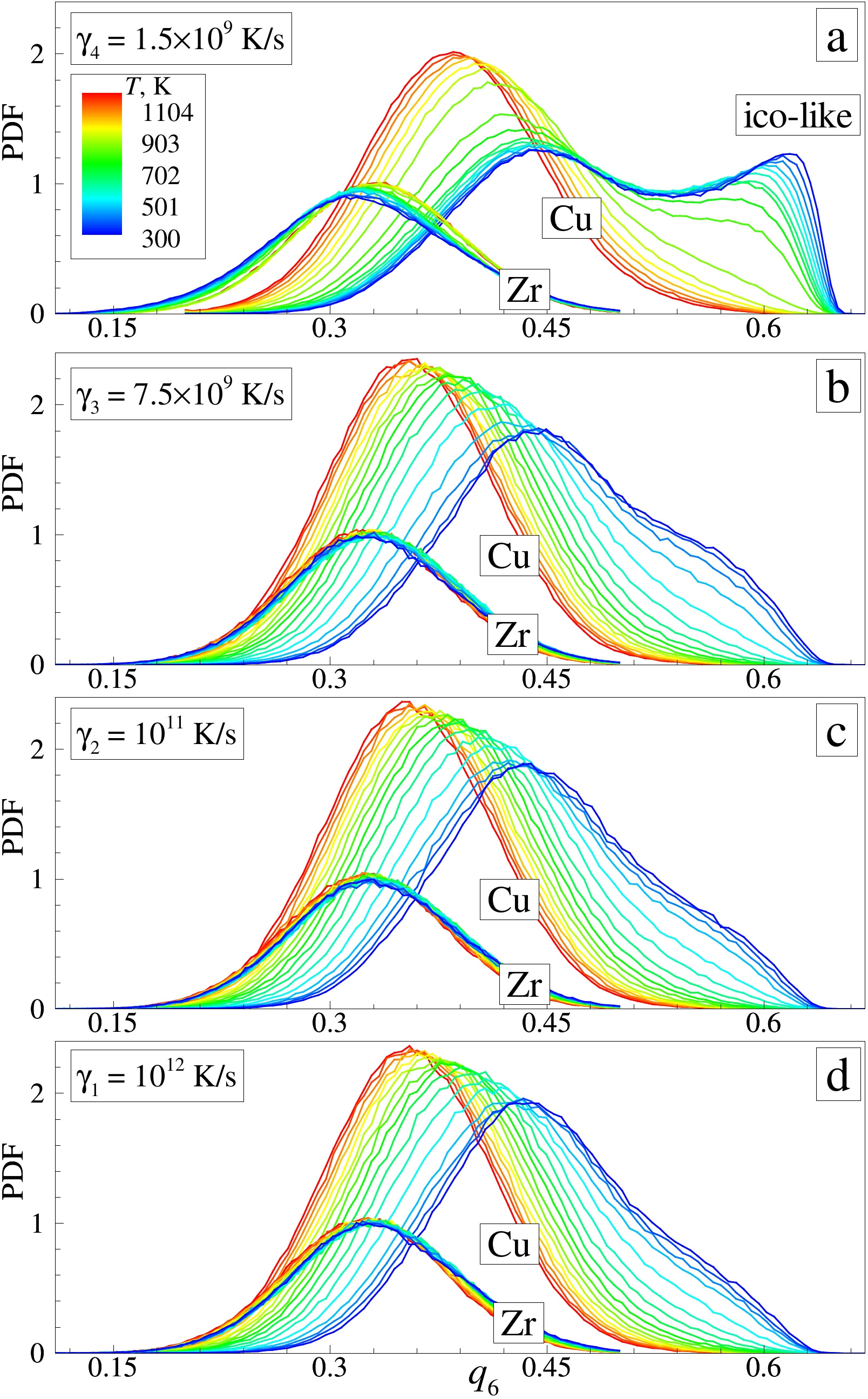}\\
\caption{(Color online). Local orientational order of the ${\rm Cu_{64.5}Zr_{35.5}}$ system at different temperatures
$T$ and cooling rates $\gamma$ (indicated on the plot) presented by the probability distribution functions (PDFс) $P(q_6)$;
the PDFs are color-coded via the temperature $T$ as shown in panel (a). Values of $q_6$ were calculated via $N_{nn}=12$ to detect close packed structures. Weak dependence of the PDFs on the $\gamma$ value is observed at $\gamma \le 10^{10}$ K/s
for all $T$; for the $\gamma = 1.5\times 10^9$ K/s the formation of ico-like Cu-centered clusters (having $q_6 > 0.6$)
is clearly seen.}
\label{fig:lo12}
\end{figure}

\begin{figure}
\centering
\includegraphics[width=\columnwidth]{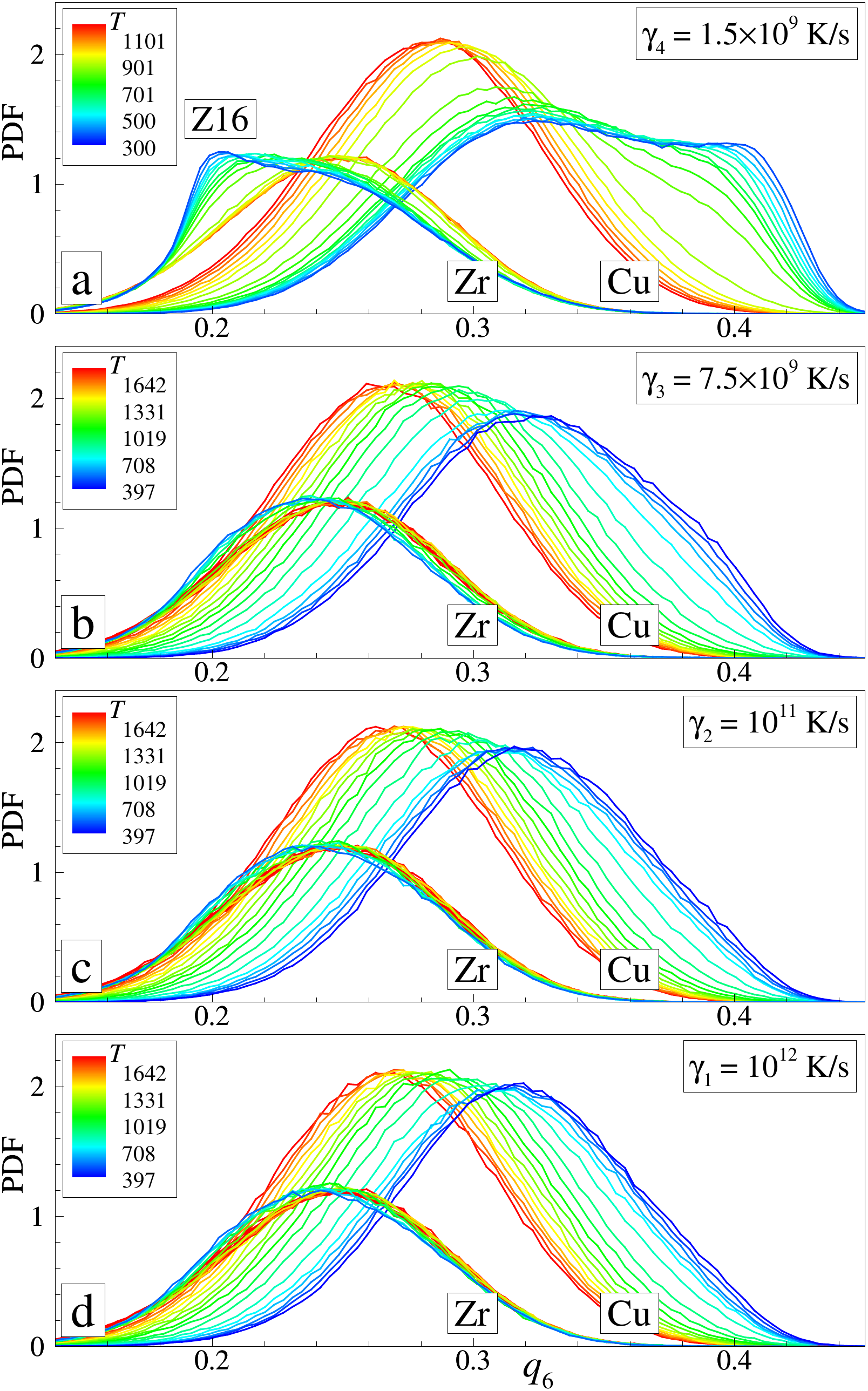}\\
\caption{(Color online). The same distributions as in Fig.~\ref{fig:lo12}; $q_6$ were calculated with using $N_{nn}=16$ to identify the Frank-Kasper phase Z16. At the $\gamma = 1.5\times 10^9$ K/s the formation of Zr-centered Z16 clusters (having $q_6 \approx 0.2$) is well pronounced.}
\label{fig:lo16}
\end{figure}

\begin{figure}
\centering
\includegraphics[width=\columnwidth]{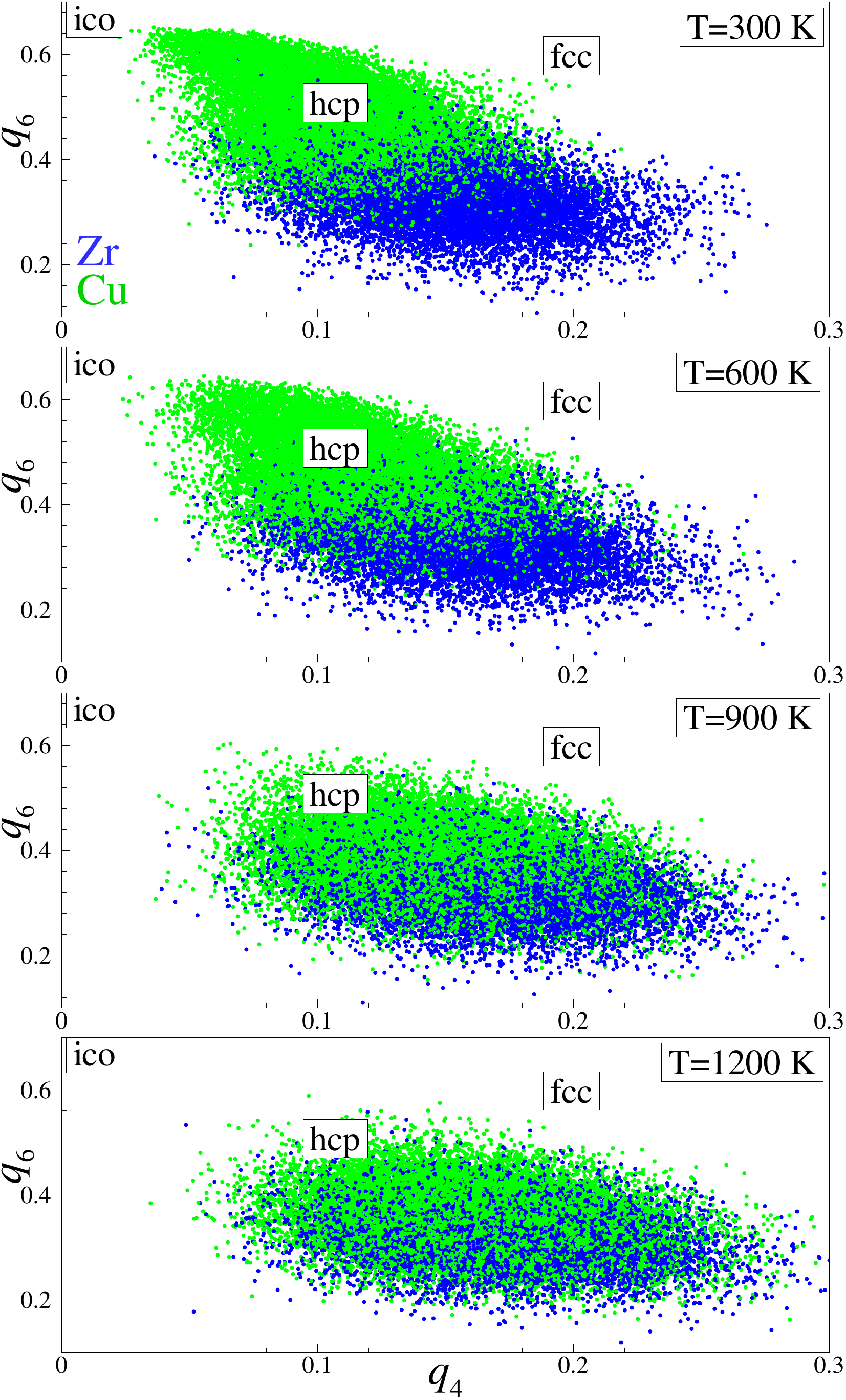}\\
\caption{(Color online). Local orientational order of the ${\rm Cu_{64.5}Zr_{35.5}}$ system on the $q_4\--q_6$ plane at different temperatures (indicated on the plot), showing the typical pathway from liquid-like to glassy state. Each point corresponds to the $(q_4,q_6)$ values for an 12 nearest neighbors cluster centered at each system atom. Green and blue points correspond to copper and zirconium atoms respectively. The formation of Cu-centered icosahedral-like clusters is clearly seen at room temperatures; zirconium atoms are nearly completely disordered (only a small part of the atoms has hcp-like symmetry) at all temperatures. BOP values for the perfect icosahedral, hcp and fcc clusters are also indicated. Cooling rate $\gamma$ is $1.5\times10^9$ K/s.}
\label{fig:lo12_2d}
\end{figure}

\begin{figure}
\centering
\includegraphics[width=\columnwidth]{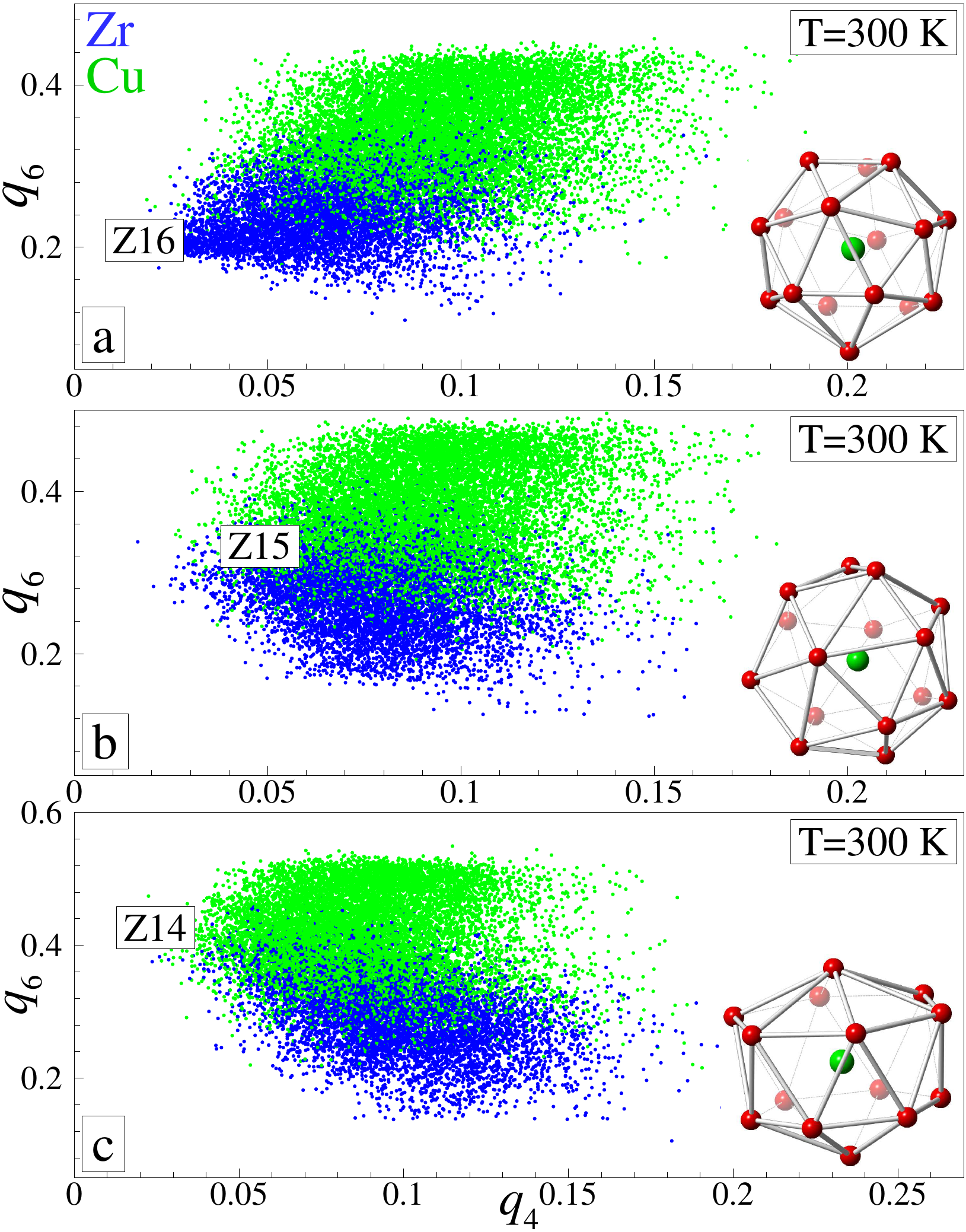}\\
\caption{(Color online). The mapping of local orientational order of room temperature glassy state of the ${\rm Cu_{64.5}Zr_{35.5}}$ system on the $q_4\--q_6$ plane. Each point corresponds to the $(q_4,q_6)$ values for an cluster with fixed value of $N_{nn}$ centered at each system atom. Bond order parameters were calculated via  $N_{nn}=16$ (panel (a)), $N_{nn}=15$ (b) and $N_{nn}=14$ (c) for both copper (green) and zirconium (blue) atoms to detect Frank-Kasper polyhedrons Z16, Z15 and Z14, respectively. BOP values for the perfect $ZN_{nn}$ clusters are also indicated. Insets show the arrangements of the neighboring atoms for the clusters under consideration \cite{Blatov2014}. Cooling rate $\gamma$ is $1.5\times10^9$ K/s.}
\label{fig4:kfs}
\end{figure}

\begin{figure}
\centering
\includegraphics[width=\columnwidth]{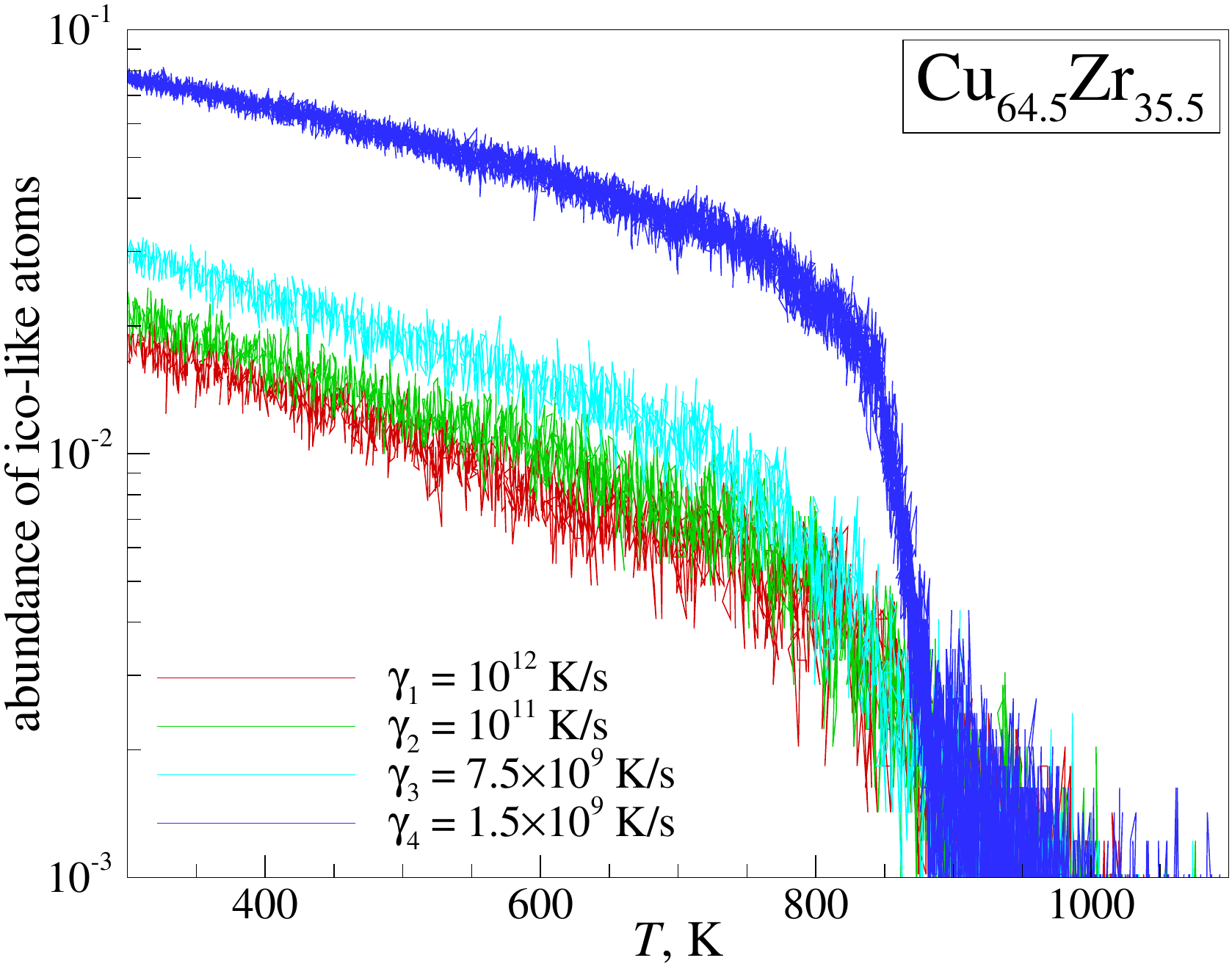}\\
\caption{(Color online). Abundance of ico-like clusters versus temperature $T$  at few $\gamma$ values (indicated on the plot). Here, atoms, having $q_6 \ge 0.6$ and $w_6 \le -0.16$ count as ico-like clusters (see, e.g. \cite{Hirata2013Science}). Ideal icosahedron has $q_6^{\rm ico}= 0.663$ and $w_6^{\rm ico} = - 0.169$ (see, Table \ref{tab1:inv}).}
\label{fig:ico}
\end{figure}

\begin{figure}
\centering
\includegraphics[width=\columnwidth]{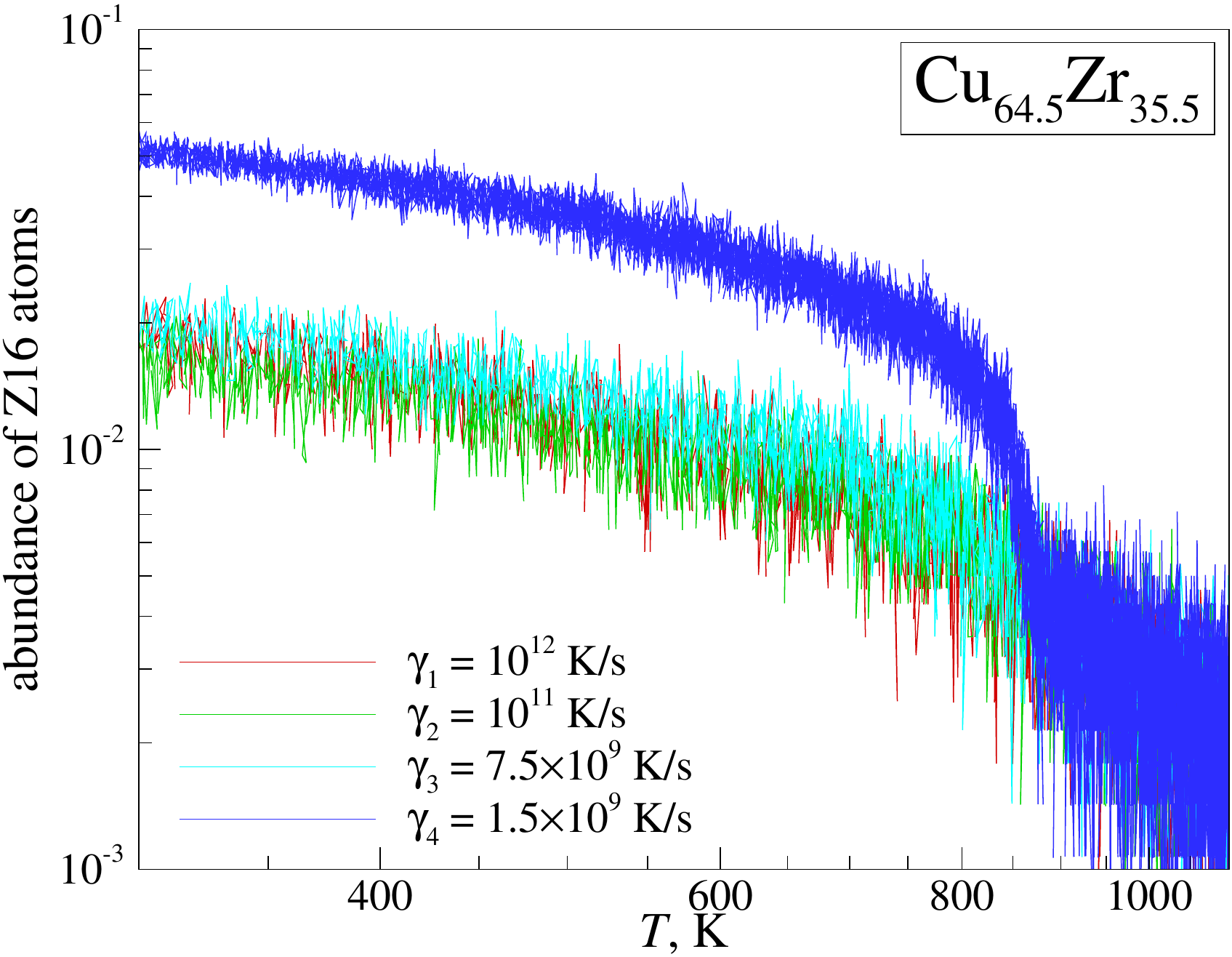}\\
\caption{(Color online). Abundance of Z16-like clusters versus temperature $T$ at few $\gamma$ values (indicated on the plot). Atoms, having  $q_4 \le 0.01$ and $|q_6 - q_6^{\rm Z16}| \le 0.01$ count as Z16-like clusters. Ideal Z16 polyhedron has $q_4^{\rm Z16}= 0.01$ and $q_6^{\rm ico} = 0.192$ (see, Table \ref{tab1:inv}).}
\label{fig:kf16}
\end{figure}

To demonstrate the influence of the cooling rate on temperature evolution of local orientational order, we show in Figs.~\ref{fig:lo12} and Figs.~\ref{fig:lo16} the probability distributions $P(q_6)$ for Cu-centered and Zr-centered clusters with 12 and 16 nearest neighbors at different temperatures and $\gamma$. Such distributions for $N_{\rm nn} = 12$ and $N_{\rm nn} = 16$ are the useful indicators to detect close packed clusters and Frank-Kasper Z16 clusters respectively.

First consider the case of $N_{\rm nn} = 12$ (Fig.~\ref{fig:lo12}). For Cu-centered clusters, a decrease of the temperature causes the shift of the $P(q_6)$ maximum towards greater $q_6$ values as well as the appearance of the shoulder in the range of $q_6$ values corresponding to icosahedral ordering (see Tab.~1). Besides, the lower the cooling rate the more pronounced the shoulder. At the lowest cooling rate $\gamma_{\rm min}=1.5\cdot 10^9$ K/s, the shoulder transforms to separate peak with the maximum at $q_6>0.6$ that means pronounced icosahedral ordering \cite{KlumovPU10,KlumovPRB11,Hirata2013Science}.

 To better illustrate the growth of icosahedral ordering under cooling, we show in Fig.~\ref{fig:lo12_2d} the mapping of system structure on two-dimensional $q_4-q_6$ plane for fixed $N_{nn}=12$. So the each point in Fig.~\ref{fig:lo12_2d} corresponds to the $(q_4,q_6)$ values for an 12 nearest neighbors cluster centered at each system atom. We see that with the decrease of temperature the $q_4-q_6$ distribution of Cu-centered clusters essentially transforms. Namely, the ``tail'' located in the $(q_4,q_6)$ range corresponding to icosahedral clusters appears. Note that the distribution of Zr-centered clusters does not demonstrate noticeable change (compared with the results for 16 neighbor clusters in Fig.~ref{fig:lo16}), remaining to be disordered (liquid-like) at all temperatures.

 In the case of $N_{\rm nn} = 16$ (Fig.~\ref{fig:lo16}),  there are no essential changes in $P(q_6)$ with temperature at cooling rates in the range $\gamma\in (10^{12}, 7.5\cdot10^{9})$ K/s; the dependence of $P(q_6)$ on cooling rate is also weak for that range. But for $\gamma=1.5\cdot10^{9}$, we see that $P(q_6)$ maximum for the Zr-centered clusters essentially shifts under cooling towards $q_6\simeq 0.2$ corresponding to ideal value for Frank-Kasper polyhedra Z16 (see Tab.1). Note that the $P(q_6)$ for Zr-centered 12 neighbour clusters demonstrate no essential change under cooling, compare Fig.~\ref{fig:lo12} and Fig.~\ref{fig:lo16}. So we conclude that noticeable amount of Frank-Kasper Z16 polyhedra is expected at only the lowest cooling rate $\gamma_{\rm min}=1.5\cdot 10^9$ K/s and for temperatures which are close to $T=300$ K.

  To better understand the structure of the system at $\gamma_{\rm min}$, we plot in the Fig.~\ref{fig4:kfs} the mapping of system structure on $(q_4, q_6)$ plane for $N_{\rm nn} = (14, 15, 16)$ at $T=300$ K, $\gamma_{\rm min}=1.5\cdot 10^9$ K/s. The pictures suggest the presence of noticeble amount of Z15 and Z16 clusters in the glassy system obtained at the lowest cooling rate. At the same time, the abundance of Z14 clusters is expected to be negligible. We suggest that the abundance of Z15 is noticeble because it can be obtained by relatively small disturbance of Z16.

 By using the BOP method, it is easily to estimate the abundance of any local clusters observing in the system. In Fig.~\ref{fig:ico} and Fig.~\ref{fig:kf16} we show the abundances of ico-like ($n_{{\rm ico}}$) and Frank-Kasper-like ($n_{{\rm FK}}$) clusters in  ${\rm Cu_{64.5}Zr_{35.5}}$ system as function of temperature at different $\gamma$ values.

 At all the cooling rates, we see drastic increase of $n_{{\rm ico}}$ at $T\lesssim T_g$ indicating the gain of icosahedral ordering approaching the glass transition; at lower cooling rates this tendency becomes more pronounced in agreement with Fig.~\ref{fig:lo12}, Fig~\ref{fig:lo12_2d}.  The temperature dependencies of the abundance of Z16 Frank-Kasper polyhedra $n_{\rm FK}(T)$ at different cooling rates also demonstrates the increase at $T\lesssim T_g$ . But cooling rate dependence of $n_{\rm FK}$ is almost negligible at all values of $\gamma$ except the $\gamma_{\rm min}$ at which the drastic increase of $n_{{\rm FK}}$ takes place.

\begin{figure}
  \centering
  \includegraphics[width=0.85\columnwidth]{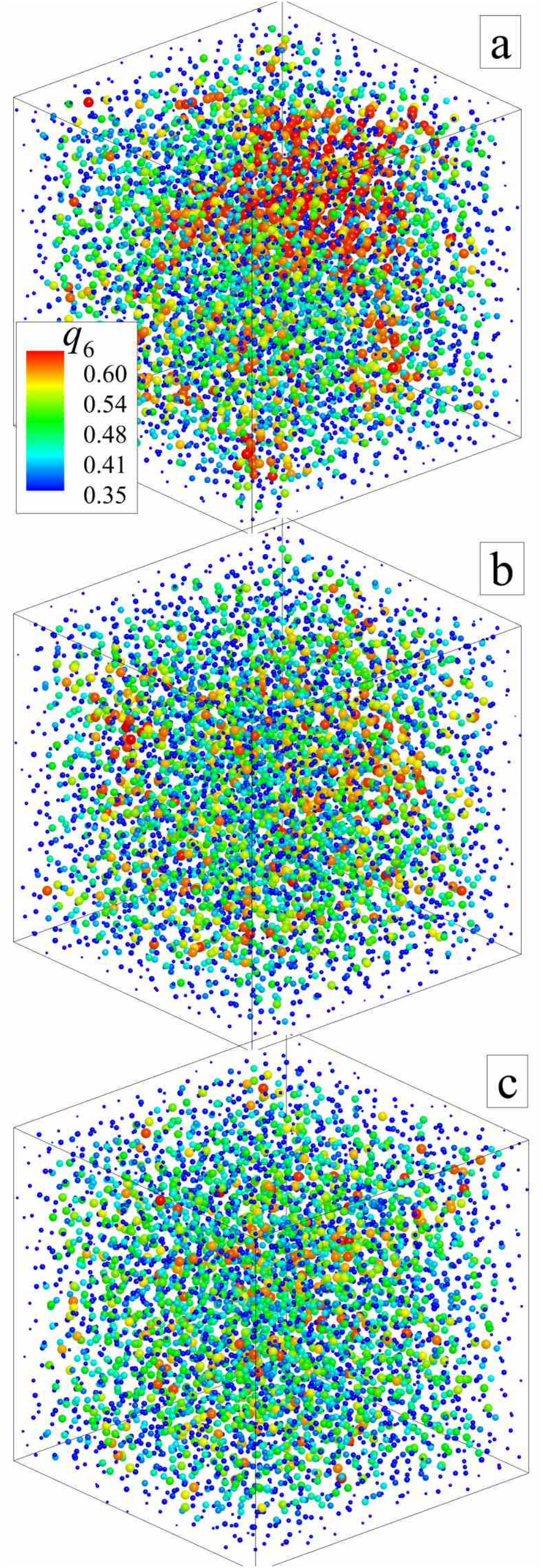}\\
  \caption{(Color online) Snapshots of the system for the final glassy state at $T=300$~K obtained at different cooling rates: (a) $\gamma=1.5\cdot 10^9$ K/s; (b) $\gamma=7.5\cdot 10^9$ K/s; (c) $\gamma=10^{11}$ K/s. Particles are colored and sized according to the value of $q_6$ for $N_{nn}=12$: so the red/orange particles with bigger size correspond to the centers of icosahedraly ordered clusters.}
  \label{fig:snapshots}
\end{figure}

 In Fig.~\ref{fig:snapshots} we show typical snapshots of the system for the final glassy state at $T=300$~K obtained at different cooling rates. Particles are colored and sized according to the value of $q_6$ for $N_{nn}=12$: so the red/orange particles with bigger size correspond to the centers of icosahedraly ordered clusters. We see that, except the increase of $n_{{\rm ico}}$, a decrease of the cooling rate causes the growth of the spatial areas where icosahedral clusters tend to connect to each other and form medium-range ``superclusters''.

So our results suggest that a decrease of the cooling rate causes essential change of the structure of ${\rm Cu_{64.5}Zr_{35.5}}$ metallic glass. In particular, the lower the cooling rate the more pronounced icosahedral ordering takes place. Note that the same conclusions has been earlier made in \cite{Zhang2015PRB_2,Yanjun2015TrNonFerMetChin}.
 As follows from the results presented above, the abundances of both the icosahedral clusters and Frank-Kasper ones increase sharply at the lowest used cooling rate $\gamma_{\rm min}=1.5\cdot 10^9$ K/s. To understand the origin of this effect, we will further study the structural properties of the system at this cooling rate in more details.

\subsection{Short- and medium-range order of glassy state at the lowest cooling rate\label{sec:lowest_v}}

 Studying the final structure of the glassy state obtained at the lowest used cooling rate $\gamma_{\rm min }=1.5\cdot 10^9$ K/s, we was surprised to see that amorphous disordered structure contains area with crystalline ordering. It can be seen in Fig.~\ref{fig:snapshots}a where spatial distribution of centers of icosahedral clusters (red/orange particles) reveals detectable area of some regular ordering.

\begin{figure}
  \centering
  \includegraphics[width=\columnwidth]{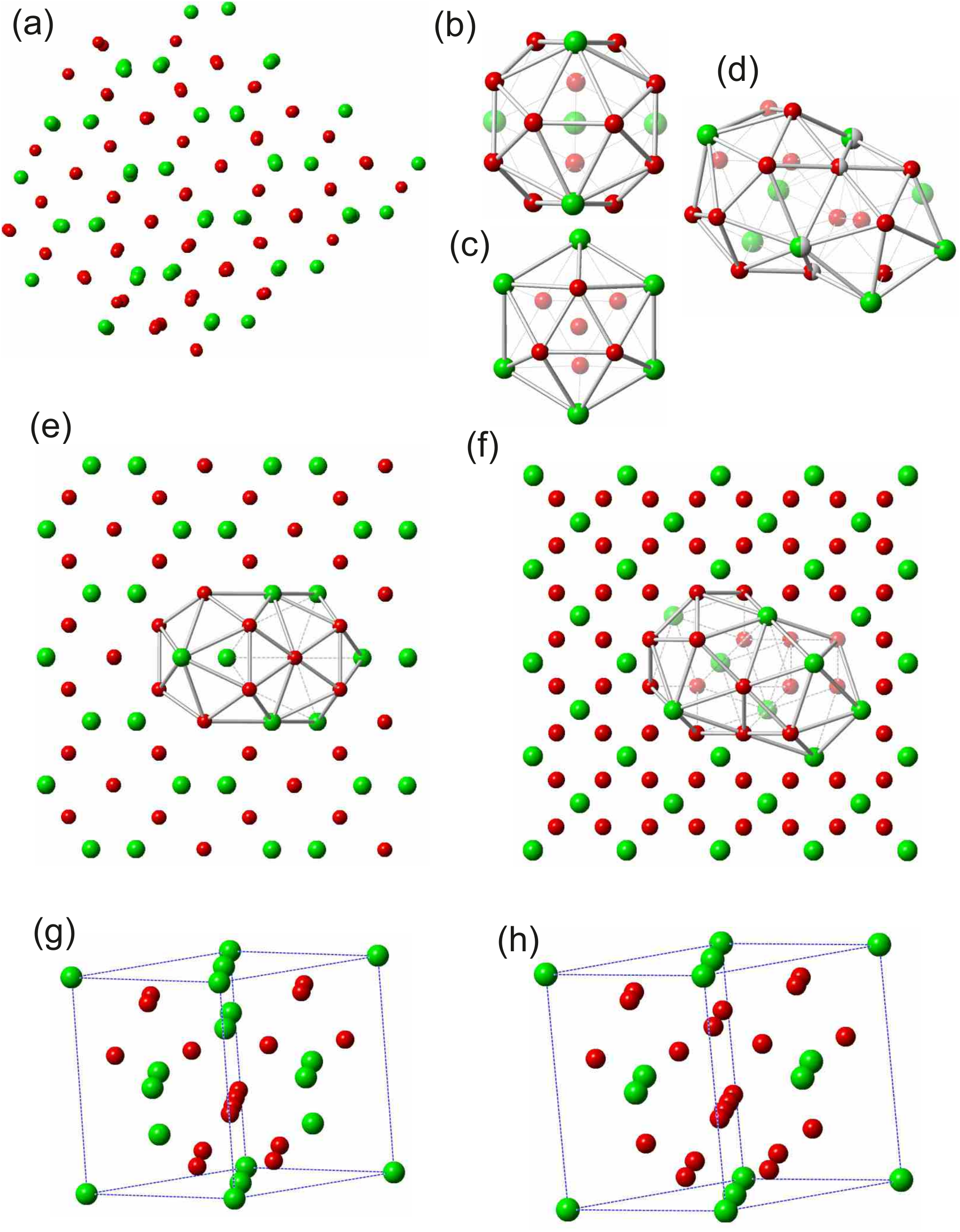}\\
  \caption{(Color online) (a) The snapshot of ${\rm Cu_2Zr}$ nanocrystal grain found in the system at cooling rate $\gamma_{\rm min }=1.5\cdot 10^9$ K/s; (b,c) The building blocks of ${\rm Cu_2Zr}$ compound: Zr-centered 17-atom Frank-Kasper polyhedron Z16 (b) and Cu-centered 13-atom icosahedral-like cluster (c); (d) The face-sharing mechanism of cluster connection. The way these cluster build in lattice of ${\rm Cu_2Zr}$ is shown in (e,f); (g) is the unit cell of ${\rm Cu_2Zr}$ compound and (h) is the unit cell of ${\rm Cu_5Zr}$ one. The pictures demonstrate isomorphism between these structures.}
  \label{fig:cu2zr}
\end{figure}

Detailed analysis of the snapshots reveals that glassy structure of the system includes nano-sized crystalline grain of ${\rm Cu_2Zr}$ compound which has the structure of ${\rm Cu_2Mg}$ Laves phase (see Fig.~\ref{fig:cu2zr}). Note that the structure of ${\rm Cu_2Zr}$ compound has not been directly observed yet neither in computer simulations nor in experiments. The stability of the compound is also an open issue (see discussion in sec.~\ref{sec:disc2}).

Examining the structure of the observed ${\rm Cu_2Zr}$ compound, one can see that it consists of two building blocks: Zr-centered 17-atom Frank-Kasper polyhedron Z16 (Fig.~\ref{fig:cu2zr}b) and Cu-centered 13-atom icosahedral-like cluster (Fig.~\ref{fig:cu2zr}c). These clusters are connected to each other by face-sharing mechanism (Fig.~\ref{fig:cu2zr}d). The way these cluster build in ${\rm Cu_2Zr}$ lattice is shown in Fig.~\ref{fig:cu2zr}(e,f).

We should also notice that observed crystal structure of ${\rm Cu_2Zr}$  is isomorphous with that for ${\rm Cu_5Zr}$ intermetallic compound (compare Fig.~\ref{fig:cu2zr}g and h). Indeed the only difference between these structures is the stoichiometry; replacing four Cu atoms in the unit cell of ${\rm Cu_5Zr}$ on Zr we get exactly the same structure as pictured in Fig.~\ref{fig:cu2zr}g.

These facts suggest that structural elements of the intermetallic compounds, both stable and metastable, may be the same as those for disordered amorphous structures \cite{Sun2016SciRep}. Under certain conditions, the local ordering of these elements may form nanocrystalline grains in amorphous structure of rapidly quenched sample. In our case, the noncrystalline grain consists of both icosahedron and Frank-Kasper polyhedron Z16 (see Fig.~\ref{fig:cu2zr}). That can explain non-monotonous cooling rate dependence of abundances of local cluster mentioned above, especially the Frank-Kasper Z16 polyhedra. Indeed, the formation of crystalline grain whose structure consists of some clusters has to lead to an increase of their amount. The results obtained suggest that the existence of significant amount of Frank-Kasper Z16 polyhedra in ${\rm Cu_{64.5}Zr_{35.5}}$ metallic glasses is related to the formation of ${\rm Cu_2Zr}$ compound. The same idea has been recently proposed in Ref.\onlinecite{Zemp2014PRB}.

\section{Discussion}

\subsection{Cooling rate dependence of glass properties\label{sec:disc1}}

Above, we have shown that properties of the ${\rm Cu_{64.5}Zr_{35.5}}$ metallic glass obtained by continuous cooling of the liquid are essentially cooling rate dependent. This result is in agreement with previous studies of Cu-Zr system \cite{Zhang2014AppPhysLett,Zhang2015PRB_2,Yanjun2015TrNonFerMetChin} as well as other glassformers \cite{Vollmayr1996JCP}. The general issue is the correspondence between simulated and experimentally measured properties of a glass which are obtained at cooling rates differing by many orders of magnitude.

The possible way to answer this question is studying cooling rate dependence of some glass property to find an empirical relation and extrapolate results to the range of $\gamma$ hardly available for direct simulations. An instructive example is the sub-$T_g$ annealing method proposed in Refs.~\onlinecite{Zhang2014AppPhysLett,Zhang2015PRB_2}. The method is based on two statements: 1) cooling rate dependence of glass potential energy obeys logarithmic law at whole range of $\gamma$; 2) annealing of the system at temperatures near the glass transition temperature $T_g$ is equivalent to effective reducing of the cooling rate. The method has been tested on Cu-Zr glass within the range of cooling rates $\gamma\in(10^{9},10^{13})$ K/s. The results suggest that starting premises of the method are reasonable \cite{Zhang2014AppPhysLett,Zhang2015PRB_2}. But our results show that situation is not so clear. Despite of similar ranges of $\gamma$ used in simulation, we see essential deviation of our $E^{\rm(300)}_{p}(\gamma)$ dependence from that obtained in \cite{Zhang2015PRB_2} at $\gamma < 10^{10}$ K/s (inset in Fig.~\ref{fig1:Epot})).  We suggest that it is due to the formation of nanocrystallite which causes the lowering of potential energy and so the deviation of $E^{\rm(300)}_{p}(\gamma)$ from logarithmic law observed in \cite{Zhang2015PRB_2} for completely amorphous systems. (Nano)crystallization is a spontaneous process which can occur at given conditions (e.g., cooling rate) with certain probability. So the authors of Refs.\onlinecite{Zhang2014AppPhysLett,Zhang2015PRB_2} may not observe it during a single simulation. The observed nanocrystallization means the ${\rm Cu_{64.5}Zr_{35.5}}$ system modeled by EAM potential by Mendelev et.al. may not be so good glass-former as the real alloy. So the additional tests of sub-$T_g$ annealing method on other model glassformers are needed.

\subsection{The structure and stability of ${\rm Cu_2Zr}$ compound\label{sec:disc2}}

Remind that, simulating the ${\rm Cu_{64.5}Zr_{35.5}}$ glass-forming alloy at cooling rate of $1.5\cdot 10^9$ K/s, we observe formation of nanocrystal of ${\rm Cu_2Zr}$ compound which has the structure of ${\rm Cu_2Mg}$ Laves phase. The results is very interesting because the structure of ${\rm Cu_2Zr}$ has not been directly observed yet.

The ${\rm Cu_2Zr}$  intermetallic compound was firstly mentioned by Kneller and co-authors in Ref. \onlinecite{Kneller1986ZMetallkund}. Studying the phase diagram of Cu-Zr system the authors observed ${\rm Cu_2Zr}$ phase as well as three other new phases: ${\rm Cu_{24}Zr_{13}}$, ${\rm CuZr_{(1+y)}}$, ${\rm Cu_5Zr_8}$. The structure of the ${\rm Cu_2Zr}$ compound was not be identified. The results of Kneller were criticized by Arias and Abriata \cite{Arias1990BullAPD} who had not observed the formation of  ${\rm Cu_2Zr}$. Later, the formation of the compound was suggested in Ref.~\onlinecite{Zaitsev2003HighTemp} on the basis of thermodynamic properties analysis using the Knudsen method of mass-spectrometry.

The stability of ${\rm Cu_2Zr}$ compound has been also predicted by ab-initio calculations \cite{Ghosh2007ActaMater, Du2014JAlloysCompd}. In the absence of experimental structural data, the structure of the compound has been chosen as that for ${\rm Au_2V}$ alloy. This issue has also been investigated by Tang and Harrowell using classical molecular dynamic simulations with EAM potential \cite{Tang2012JPCM}. Based on the analogy with Lennard-Jones binary alloy with similar size ratios \cite{Pedersen2010PRL}, the authors propose that the structure of the ${\rm Cu_2Zr}$ compound is ${\rm MgZn_2}$-type Laves phase. IT has been shown that this structure can be stable in certain temperature-concentration domain.

Despite of contradictory data, the ${\rm Cu_2Zr}$ compound was included as one of the possible phase for the thermodynamic assessment of phase diagram of the Cu-Zr system  \cite{Wang2006Calthad}. The results of the assessments show the stability range of the compound similar to those obtained experimentally in Ref. \onlinecite{Kneller1986ZMetallkund}.

So we conclude the structure of ${\rm Cu_2Zr}$ compound is still the matter of debates. We have directly observed self-assembling of the compound for the first time and so our results suggest its possible structure. The experimental validation of that structure is the matter of future investigations.

Another issue is the stability of the compound in the EAM model of Cu-Zr system. We firstly found formation of the nanocrystalline grain during fast cooling and then showed that the same structure formed during isothermal annealing (see Fig.~\ref{fig1:Epot}b). It is important that annealing at any temperature in the range of $(700, 900)$ K leads to similar structure with nano-sized grain of ${\rm Cu_2Zr}$ compound. That means the crystal nucleus forming in the system under the given conditions does not reach the critical value and so does not growth indefinitely. The questions arise: is this nucleus stable at very long annealing times? is the growth of the nucleus limited by local stresses imposed by periodic boundary conditions? To answer these issues the large-scale simulations requiring several months of supercomputer time are needed.

\section{Conclusions}

Doing molecular dynamics simulations, we show that structural evolution of ${\rm Cu_{64.5}Zr_{35.5}}$ glass-forming alloy essentially depends on cooling rate. In particular, a decrease of the cooling rate leads to a increase of abundances of both the icosahedral-like clusters and Frank-Kasper Z16 polyhedra. The amounts of these clusters in the glassy state drastically increase at the $\gamma_{\rm min}=1.5\cdot 10^{9}$ K/s. It is explained by the formation of nano-sized crystalline grain of ${\rm Cu_2Zr}$ compounds whose structure consists of these clusters. Note that the structure of ${\rm Cu_2Zr}$ compound is directly observed for the first time and so this results may stimulate the further investigations to its experimental validation.

We also show that the formation of ${\rm Cu_2Zr}$ compound is also observed during the isothermal annealing of the system at temperatures near the glass transition temperature. That means the ${\rm Cu_{64.5}Zr_{35.5}}$ alloy modeled by EAM potential may not be so good glassformer. Probably, it will totally crystalize at longer annealing and/or large system size.

\section{Acknowledgments}
Molecular dynamics simulations and the access to "Uran" cluster were supported by the Russian Science Foundation (grant RNF №14-13-00676).
Structural analysis was supported by Russian Science Foundation (grant Nr. 14-12-01185). Program codes for local orientational order analysis were developed within the frames of the grant RNF 14-50-00124. BAK was partially supported by the A*MIDEX grant (Nr. ANR-11-IDEX-0001-02) funded by the French Government “Investissements d’Avenir” program.

\bibliography{our_bib_cu_zr}
\end{document}